\documentclass[preprint,showpacs,preprintnumbers,amsmath,amssymb]{revtex4}

\usepackage{graphicx}

\begin{document}

\title{
Possible latitude effects of Chern-Simons gravity on quantum interference 
}
\author{Hiroki Okawara}
\author{Kei Yamada}
\author{Hideki Asada} 
\affiliation{
Faculty of Science and Technology, Hirosaki University,
Hirosaki 036-8561, Japan} 

\date{\today}

\begin{abstract}
It has been recently suggested that, 
as a gravitational Aharonov-Bohm effect 
due to a gravitomagnetic potential, 
possible effects of Chern-Simons gravity 
on a quantum interferometer
are dependent on 
the latitude and direction of the interferometer on Earth 
in orbital motion around Sun. 
Continuing work initiated in the earlier publication 
[Okawara, Yamada and Asada, Phys. Rev. Lett. 109, 231101 (2012)], 
we perform numerical calculations of time variation 
in the induced phase shifts for nonequatorial cases. 
We show that the maximum phase shift at any latitude 
might occur at 6, 0 (and 12), and 18 hours (in local time) of each day, 
when the normal vector to the interferometer is 
vertical, eastbound and northbound, 
respectively.
If two identical interferometers were located at different latitudes, 
the difference between two phase shifts that are measured 
at the same local time would be $O(\sin \delta\varphi)$ 
for a small latitude difference $\delta\varphi$. 
It might thus become maximally $\sim 20$ percents 
for $\delta\varphi \sim 10$ degrees, for instance. 
\end{abstract}

\pacs{04.25.Nx, 04.50.-h, 04.80.Cc}

\maketitle

\section{Introduction}
As a fundamental problem, 
the interplay between the quantum theory and 
the gravitational physics has been studied 
mostly by theoretical experiments \cite{Refs}. 
Corella, Overhauser, and Werner (COW) \cite{COW}
succeeded in a first experiment involving 
both the Plank constant $h$ and the gravitational constant $G$ 
by using a neutron interferometer.  
In the COW experiments, a neutron interferometer is tilted, 
such that a neutron beam path I is higher above the surface of 
Earth than the other path segment II, 
causing a gravitationally induced phase shift of the neutron 
de Broglie waves on path II relative to path I. 
The gravitationally induced phase shift was experimentally 
observed \cite{COW, Sakurai-Book}. 
This experimental result means that the classical gravitational field 
(at the Newtonian order) affects a quantum particle 
as well as a classical one. 
In recent decades, there has been technological progress 
in quantum experiments including neutron interferometers 
and quantum optics. 
As a next step, 
current attempts to probe general relativistic effects 
in quantum mechanics focus on 
precision measurements of phase shifts in 
quantum interferometers (e.g. \cite{Zych}). 
Hogan has recently proposed an ambitious idea 
to use quantum interferometers as an experimental probe 
of a quantum spacetime at the Planck scale \cite{Hogan}. 
Quantum experiments may play a role in probing 
an intermediate regime between general relativistic gravity 
and Planck scale physics. 

In addition to the above fundamental issue, 
current astronomical observations, such as the apparent 
accelerated expansion of the Universe, suggest 
a possible infrared modification to general relativity. 
For instance, dark energy is introduced to explain 
the observed accelerated expansion 
by means of an additional energy-momentum component 
in the right-hand side of the Einstein equation. 
As an alternative approach for interpreting 
the present accelerated expansion, 
the left-hand side of the Einstein equation, 
equivalently the Einstein-Hilbert action, 
could be modified in various ways (nonlinear curvature terms, 
higher dimensions, and so on). 
The Chern-Simons (CS) correction is one of modified gravity models. 
The CS modification is not an {\it ad hoc} extension, 
but it is actually motivated by both string theory, as a
necessary anomaly-canceling term to conserve unitarity
\cite{Polchinski}, 
and 
loop quantum gravity (LQG), as 
a counter term for the anomaly\cite{Ashtekar} 
and recently as the emergence of the CS gravity when 
the Barbero-Immirzi parameter of LQG is promoted to 
a scalar field and the Holst action is coupled to fermions 
\cite{Taveras}. 
Alexander and Yunes have recently pointed out that 
CS gravity possesses the same parametrized post-Newton (PPN) 
parameters as general relativity, 
except for the inclusion of a new term, proportional to the CS coupling 
and the curl of the PPN vector potential \cite{AY1, AY2}. 
They have also shown 
that this new correction 
might be used in {\it classical} experiments, 
such as the Gravity Probe B (GPB), 
to bound CS gravity 
(see \cite{AY2009} for an extensive review of CS modified gravity). 
Their proposal has been implemented by Ali-Haimoud and Chen 
\cite{Ali-Haimoud} to constrain CS gravity. 
Dyda, Flanagan and Kamionkowski \cite{Dyda} 
have constrained nondynamical CS gravity 
(or equivalently CS gravity with a canonical background 
cosmological scalar) by studying vacuum instabilities. 

It is interesting to study, 
as an attempt to probe quantum gravity, 
possible effects of the CS modified gravity on {\it quantum} experiments. 
Along this course, 
Nandi and his collaborators \cite{Nandi} 
have discussed the quantum phase shift in a CS modified gravity model, 
where an isolated gravitating body was considered. 
Their conclusion is that 
the induced shifts by the spin of the body 
are too tiny to be observed. 
However, Earth's orbital angular momentum 
($\sim 3 \times 10^{40} \; \mbox{kg} \cdot \mbox{m}^2 \mbox{s}^{-1}$) 
is much larger than its spin angular momentum 
($\sim 7 \times 10^{33} \; \mbox{kg} \cdot \mbox{m}^2 \mbox{s}^{-1}$). 
Both of the axial vectors may play a role in CS gravity. 
Therefore, the present authors \cite{OYA} have considered 
gravitationally interacting bodies 
in order to investigate the quantum mechanical effects of 
Earth's orbital angular momentum in CS gravity. 
It has been suggested that a CS modified gravity theory 
may predict daily and seasonal phase shifts in quantum interferometers, 
which are in principle distinct from the general relativistic effects. 
This feature can be currently used as a quantum tool 
to probe CS gravity. 

The numerical calculation in the previous paper \cite{OYA} 
has been done only for the equatorial case for its simplicity. 
In the present paper, therefore, we shall perform numerical calculations of 
possible daily and seasonal variations 
in quantum interference at nonequatorial locations. 
This might be important for high-precision 
quantum experiments, because of the geographical reason. 
Namely, almost all quantum interferometers are located not 
at the equator of Earth but at the middle latitude region 
between $23^{\circ}26^{'}22{''}$ and $66^{\circ}33^{'}39{''}$, 
including most areas of the USA, many European countries 
and Japan. 
We shall argue that the CS latitude effects are important 
in experiments.  
If some variation were marginally detected in the future 
(presumably at a low signal-to-noise ratio),  
it would be difficult to disentangle the CS signal from 
other effects without taking account of the latitude effect. 
Comparing phase measurements at two (or more) latitudes 
would be helpful for improving the CS bound or 
distinguishing the CS signal from others. 
Namely, a signal-to-noise ratio could be increased 
by a combined analysis of phase measurements at different latitudes.

\section{Quantum interference induced by CS gravity}
In this section, 
we summarize the basics of computing 
the quantum phase shift induced by CS gravity, 
following Ref. \cite{OYA}. 

\subsection{CS modified gravity}
CS gravity modifies general relativity 
via the addition of a correction to 
the Einstein-Hilbert action, namely \cite{Jackiw, Guarrera}
\begin{eqnarray}
S_{\mbox{CS}} \equiv \frac{1}{16\pi G}\int d^4x\frac{1}{4}fR^{\star}R , 
\label{CS-action}
\end{eqnarray}
where 
$f$ is a prescribed external field (with units of area in 
geometrized units), 
$R$ is the Ricci scalar, and the star stands for the dual operation, 
such that 
\begin{eqnarray}
R^{\star} R 
\equiv \frac12 R_{\alpha\beta\gamma\delta} 
\epsilon^{\alpha\beta\mu\nu}
R^{\gamma\delta}{}_{\mu\nu} , 
\end{eqnarray}
with $\epsilon_{\mu\nu\delta\gamma}$ 
the totally-antisymmetric Levi-Civita tensor and 
$R_{\mu\nu\delta\gamma}$ the Riemann tensor.
We concentrate on a nondynamical (kinematical) model of 
CS modified gravity, where we assume that $f$ 
does not have dynamics (no kinetic term). 
We focus on a slowly changing $f$ where $\dot f$ is considered but 
no second-time derivatives of $f$ appear. 
The nondynamical CS theory is tractable 
and could become a good approximation in weak fields, 
though there remains a possible evolution problem 
of the external field $f$ 
(presumably near the central region) consistent with 
the Pontryagin constraint 
and recent studies suggest that the nondynamical CS gravity model 
is an overconstrained system of equations 
\cite{Grumiller,Sopuerta}. 
A full dynamical study of seeking approximate solutions 
for rotating extended bodies has yet to be carried out 
\cite{AY2009}. 
It has been accomplished in the slow-rotation approximation 
\cite{rotation}.

The weak-field solution to the CS modified field equations 
in PPN gauge is given by \cite{AY1, AY2, AY2009}
\begin{eqnarray}
g_{00}&=& -1 + 2U -2 U^2 + 4 \Phi_1 + 4 \Phi_2 + 2 \Phi_3 
+ 6 \Phi_4 + O(6) , 
\\
g_{0i}&=& -\frac72 V_i - \frac12 W_i +2 \dot f (\nabla\times V)_i 
+O(5), 
\label{g0i}
\\
g_{ij}&=& (1 + 2U) \delta_{ij} + O(4) , 
\end{eqnarray}
where ${U, \Phi_1, \Phi_2, \Phi_3, \Phi_4, V_i, W_i}$ 
are PPN potentials (e.g. \cite{Will}), 
$O(A)$ stands for PN remainders of order $O(1/c^A)$ 
for the light speed $c$ 
and the dot denotes the derivative with respect to $x^0 \equiv ct$. 

Henceforth, we investigate the quantum mechanical effects 
at the linear order of the $\dot f$ term in Eq. (\ref{g0i}). 
See \cite{OYA} for discussions 
on the second (or higher) order contributions. 

Following \cite{AY1}, 
let us consider a system of nearly spherical bodies 
in the standard PPN point-particle approximation, 
since we concentrate on weakly-gravitating bodies.  
For the above vector potential $V_i$, 
the CS correction to the metric becomes 
in the barycenter frame 
\cite{AY1, AY2, AYcorr, AY3}
\begin{eqnarray}
\delta_{\mbox{CS}} g_{0i}=
\frac{2G}{c^3} \sum_A \frac{\dot f}{r_A} 
\left[\frac{m_A}{r_A}(\vec v_A \times \vec n_A)^i -\frac{J^i_A}{2r^2_A} 
+\frac{3}{2}\frac{(\vec J_A \cdot \vec n_A)}{r^2_A}n^i_A \right] ,
\label{deltag0i}
\end{eqnarray}
with $m_A$ the mass of the $A$th body, 
$r_A$ the field point distance to the $A$th body, 
$n_A^i = x_A^i/r_A$ a unit vector pointing to the $A$th body, 
$v_A^i$ the velocity of the $A$th body, 
$J_A^i$ the spin-angular momentum of the $A$th body,  
and the $\cdot$ and $\times$ operators are the flat-space 
inner and outer products. 
Note that the CS correction couples  
with the spin 
and the orbital angular momenta. 

\begin{figure}[t]
\includegraphics[width=12cm]{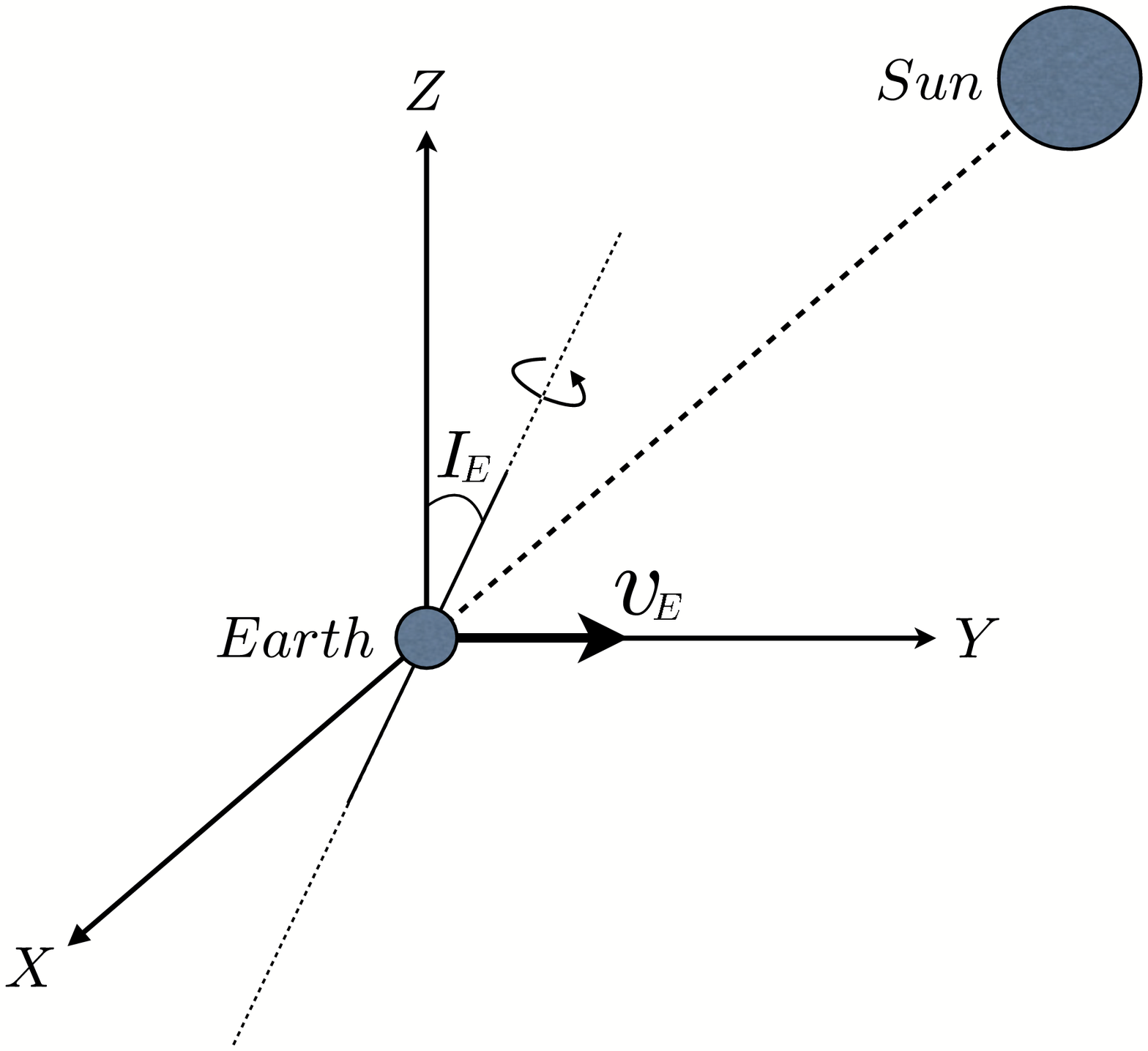}
\includegraphics[width=12cm]{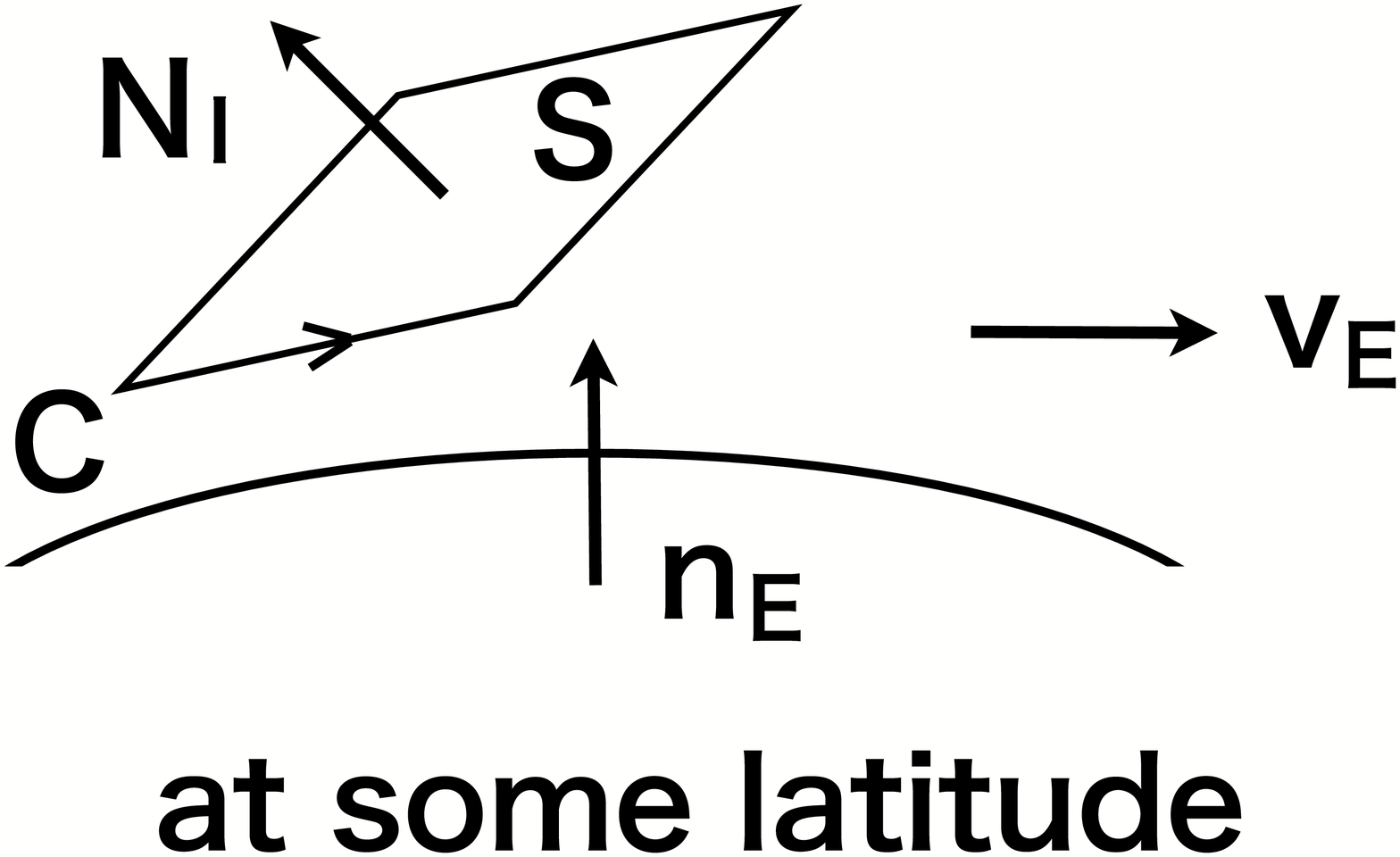}
\caption{ 
Quantum interferometer on Earth orbiting Sun. 
The orbital plane is chosen as the $X$-$Y$ plane. 
Earth's axis and orbital velocity are 
denoted by $I_E$ and $v_E$, respectively. 
Top: Earth orbiting the barycenter. 
Bottom: Interferometer  
at a certain time and place on Earth. 
The latitude and the longitude are specified by 
the unit normal $\vec n_E$, 
which rotates in an inertial frame 
around Earth's axis 
and hence its direction changes owing to 
the orbital motion of Earth. 
In an inertial frame, the interferometer's direction  
$\vec N_I$ also 
changes as Earth rotates. 
}
\label{f1}
\end{figure}

\subsection{Quantum phase shifts in a curved spacetime} 
We consider a quantum interferometer 
that consists of a closed path $C$ (its area $S$) 
on Earth, 
as shown by Fig \ref{f1}. 

The Hamiltonian for a quantum particle in a curved spacetime 
involves $g_{\mu\nu}$. 
The linear-order correction to the Hamiltonian by $g_{0i}$ becomes 
$\delta H = mc g_{0i} v^i$ 
for a slowly moving particle \cite{LL}. 
A phase difference induced by $g_{0i}$ is thus 
expressed as \cite{Sakurai-Book}
\begin{eqnarray}
\Delta&=&
\frac{1}{\hbar} \oint_C \delta H dt 
\nonumber\\ 
&=& 
\frac{mc}{\hbar} \oint_C \vec g \cdot d\vec r , 
\label{Delta1}
\end{eqnarray}
where $\vec g$ denotes $(g_{01}, g_{02}, g_{03})$, 
$m$ denotes the quantum particle mass, 
$\hbar \equiv h/2\pi$ denotes Dirac's constant. 
By using Stokes theorem, $\Delta$ is rewritten in the surface 
integral form over $S$ as 
\begin{eqnarray}
\Delta=\frac{mc}{\hbar}\int_S (\vec \nabla \times \vec g) 
\cdot d\vec S . 
\label{Delta2}
\end{eqnarray}
This form has the exact analogy in the Aharonov-Bohm (AB) effect, 
because $\vec{g}$ is the gravitomagnetic potential 
in the gravito-electromagnetic description. 
The original AB effect in the phase shift, 
which was confirmed experimentally 
\cite{Chambers}, is  
$\propto \oint_C \vec A \cdot d\vec r = 
\int_S (\vec\nabla \times \vec A)\cdot d\vec S$ 
for a vector potential $\vec A$ in the electromagnetism.  
Note that 
the phase difference $\Delta$ in Eq. (\ref{Delta2}) 
is caused by 
frame-dragging effects in rotational spacetimes 
and hence it does not depend on de Broglie wavelength $\lambda$, 
in contrast to COW experiments. 

\subsection{Phase shifts induced by CS gravity} 

Let us substitute the CS term  
of Eq. (\ref{deltag0i}) 
into Eq. (\ref{Delta2}) to obtain $\Delta$ for CS gravity. 
By using an identity $\epsilon^{ijk} (1/r)_{, jkl} = 0$ 
with the Levi-Civita symbol $\epsilon^{ijk}$, 
one can see that 
the $J$-dependent part of the metric in Eq. (\ref{deltag0i}) 
always vanishes in Eq. (\ref{Delta2}), 
whereas the $v$-dependent part makes contributions. 

Since $\Delta$ involves the curl operation 
on the surface of Earth 
and Earth's radius $r_E$ is much shorter than 1AU, 
the terms associated with 
the solar mass $M_{\odot}$ in Eq. (\ref{Delta2}) 
are $O(M_{\odot} M_E^{-1} r_E^3 1\mbox{AU}^{-3}) \sim 10^{-9}$ 
smaller than those associated with Earth's mass $M_E$, 
so that the terms  
with 
the solar mass (and other planetary ones) 
can be safely neglected. 
Henceforth, we focus on Earth's mass 
(also its spin and orbital angular momentum) 
in CS gravity. 
Hence, Eq. (\ref{Delta2}) becomes \cite{OYA}
\begin{eqnarray}
\Delta_{\mbox{CS}}&=&
\frac{2m}{\hbar c^2}\int_S \dot f\frac{GM_E}{r^3} 
\left[3(\vec v_E \cdot \vec n_E)\vec n_E -\vec v_E\right] 
\cdot\vec N_I dS 
+ O(\dot f^2)
\nonumber\\
&=&2\dot f\frac{mGM_ES}{\hbar c^2 r_E^3} 
\left[3(\vec v_E \cdot \vec n_E)\vec n_E -\vec v_E \right] 
\cdot\vec N_I 
+ O(\dot f^2) , 
\label{Delta-CS}
\end{eqnarray}
where we used $r_E \gg \sqrt{S} $ 
(Earth's radius is much larger than 
the interferometer arm length) 
and hence $r = r_E$ in the integrand. 
Here, in an inertial frame, 
$\vec v_E$ denotes Earth's orbital velocity, 
$\vec n_E$ stands for the unit vertical vector on the ground 
(at a certain latitude), 
$\vec N_I$ means the unit normal to the interferometer plane 
(See also Fig. \ref{f1}). 
The unit normal vectors $\vec n_E$ and $\vec N_I$ 
in an inertial frame change with time 
as Earth rotates. 
The change rate depends on the latitude. 
Moreover, $\vec N_I$ depends also 
on the interferometer's 
direction such as horizontal and vertical. 
In contrast to COW experiments, 
the interferometer direction 
such as North and East does matter in CS gravity. 
Therefore, the factor 
$\left[3(\vec v_E \cdot \vec n_E)\vec n_E -\vec v_E \right] 
\cdot\vec N_I$ in Eq. (\ref{Delta-CS}), 
depending on the latitude and direction, 
changes with Earth's spin and orbital motion. 
The directional dependence of the general relativistic 
phase shifts (e.g. \cite{Kuroiwa,Wajima}) 
is in principle distinct from 
that of CS gravity \cite{OYA}. 

The order-of-magnitude estimation of $\Delta_{\mbox{CS}}$ 
has been made in Ref. \cite{OYA}. 
The magnitude of Eq. (\ref{Delta-CS}) is factored as 
\begin{equation}
|\Delta_{\mbox{CS}}| \sim 4 \left(\frac{mc^2}{\hbar}\right) 
\left(
\frac{\dot f}{c} 
\frac{GM_E}{c^2r_E}
\frac{v_E}{c}\right) 
\left(\frac{S}{r_E^2}\right) , 
\label{Delta-CS2}
\end{equation}
where $[3(\vec v_E \cdot \vec n_E)\vec n_E -\vec v_E] \cdot\vec N_I 
\sim 2 v_E$. 
It is worthwhile to mention that the first fraction  
in the right-hand side of Eq. (\ref{Delta-CS2}) is 
due to the quantum mechanical physics 
and it is large enough $\sim 10^{24} \, \mbox{s}^{-1}$ 
to compensate for the factor in the second parentheses 
due to the CS gravitational effect 
$\sim \dot f c^{-1} \times 10^{-14}$, 
where $m$ is neutron mass. 
The last factor in Eq. (\ref{Delta-CS2}) 
is the squared ratio of the interferometer 
arm length (often $\sim 60 \, \mbox{cm}$ \cite{Book}) 
to Earth's radius. 
In total, the magnitude of $\Delta_{\mbox{CS}}$ is 
\begin{equation}
|\Delta_{\mbox{CS}}| \sim 10^{-3} \mbox{s}^{-1} \times 
\left(\frac{mc^2}{1 \mbox{GeV}}\right) 
\left(\frac{\dot f}{c}\right) 
\left(\frac{S}{0.4 \mbox{m}^2}\right) . 
\label{Delta-CS3}
\end{equation}

The current bound on the $\dot f$ parameter by neutron interferometry 
and a possible improvement have been discussed \cite{OYA}. 
Current measurements of phase shifts in neutron interferometry 
do not report any anomalous (daily nor seasonal) variations  
with phase measurement accuracy 
at $O(10^{-3})$. 
Current neutron interferometry, 
therefore, places a bound on CS gravity as 
$\dot f c^{-1} < 10^{0} \mbox{s}$ ($\dot f < 10^{5} \mbox{km}$), 
which is worse by three digits than the constraint 
$\dot f c^{-1} < 10^{-3} \mbox{s}$ 
by the classical experiment GPB (Gravity Probe B) 
\cite{AY1, GPB}
and also LAGEOS \cite{Smith}. 

Future progress in quantum technology 
may improve the bound. 
A bound comparable to the GPB limit would be placed, 
if neutron interferometry were sufficiently improved for 
$\Delta \times S^{-1}$ (nearly by three digits). 
For instance, $\sim 5$ meters arm length 
and $\sim 10^{-4}$ phase measurement accuracy 
are preferred. 

Werner and his collaborators have already obtained the result 
for the measured phase shift with a one-sigma statistical error bar 
as $\pm 0.34$ mrad $\sim O(10^{-4})$, where approximately, 
a total of 500 000 000 neutrons were 
counted in the interferograms over a period of 2 years 
(See \cite{Werner} for a review of observations of Aharonov-Bohm 
effects by neutron interferometry). 
Furthermore, Seki and his collaborators have recently 
developed a multilayer cold-neutron interferometer 
and experimentally the phase measurement accuracy of $0.01$ rad, 
where they used only $1.5 \times 10^5$ neutrons in a short time 
$\sim 49$ hours \cite{Seki}. 
Motivated by quantum computations, for instance, 
Pushin and his collaborators 
have demonstrated experimentally how quantum-error-correcting codes 
may be used to improve experimental designs of quantum devices 
to achieve noise suppression in neutron interferometry \cite{Pushin}.

Experimental setups usually suffer from many other seasonal variations. 
Lacking a signal, a constraint may be placed on $\dot f$. 
In the presence of a signal, on the other hand, 
one would have to eliminate all other possible sources of 
seasonal variability.

\section{Numerical Calculations} 
\subsection{Calculations in the laboratory frame} 
The time dependence of $\Delta_{\mbox{CS}}$ 
comes from the factor 
$\left[3(\vec v_E \cdot \vec n_E)\vec n_E -\vec v_E \right] 
\cdot\vec N_I$ in Eq. (\ref{Delta-CS}). 
For computing the time variation, 
we take account of Earth's parameters 
such as the inclination angle of Earth's axis 
$I_E$, 
the mean orbital angular velocity 
$\Omega_E$, 
the spin rate  
$\omega_E$, 
whereas the eccentricity of Earth's orbit has 
a tiny input to be ignored in this paper.

By straightforward calculations, 
the above factor is rearranged as 
\begin{eqnarray}
\left[3(\vec v_E \cdot \vec n_E)\vec n_E -\vec v_E \right] 
\cdot\vec N_I 
&=& ({\cal R}^{-1} \vec v_E)^{T} 
\left[3(\vec n_{E0} \cdot \vec N_{I0})\vec n_{E0} -\vec N_{I0} \right] . 
\label{factor}
\end{eqnarray}
Here, the superscript $T$ denotes 
the transposition of the vector (and matrix),  
the subscript $0$ denotes the initial time 
that is chosen as the midnight on the winter solstice day 
and ${\cal R}^{-1}$ denotes symbolically the inverse matrix of 
the product of rotational matrices, 
each of which corresponds to 
the spin of Earth, 
the inclination of Earth's rotation axis, 
its orbital rotation around the barycenter 
and the latitude angle $\varphi$, respectively.  
To obtain Eq. (\ref{factor}), we have used 
${\cal R}^{T} = {\cal R}^{-1}$ and 
${\vec a}^T {\cal R} \vec b = ({\cal R}^{-1} \vec a)^T \vec b$ 
for vectors $\vec a$ and $\vec b$. 

Equation (\ref{factor}) seems convenient, 
since $\vec N_{I0}$ is fixed in the laboratory frame 
and the combination of 
$[3(\vec n_{E0} \cdot \vec N_{I0})\vec n_{E0} -\vec N_{I0}]$ 
is thus a constant vector (as usually taken for granted 
in local experiments). 
On the other hand, 
$[3(\vec n_{E} \cdot \vec N_{I})\vec n_{E} -\vec N_{I}]$
changes with time in inertial frames 
such as the barycenter frame. 
For discussing the time evolution of Eq. ($\ref{factor}$), 
it is sufficient to consider only the rotational matrix part.

For treating the vector components in numerical calculations, 
we adopt the Cartesian coordinates $(x, y, z)$ 
in the laboratory frame associated with the quantum interferometer, 
such that the $x$, $y$ and $z$ axes can be 
along the East, North and vertical upward directions, respectively. 
Hence, $\vec n_E = (0, 0, 1)$.  
In the laboratory frame, the Cartesian components 
of the term ${\cal R}^{-1} \vec v_E$ become 
\begin{align}
({\cal R}^{-1} \vec v_E)_x 
=& 
v_E \left(
[\sin^2(\Omega_E t)\cos(I_E)+\cos^2(\Omega_E t)]\cos(\omega_E t) 
\right.
\notag\\
&
\left.
~~~-[\sin(\Omega_E t)\cos(\Omega_E t)(1-\cos(I_E))]\sin(\omega_E t)
\right) , 
\nonumber\\
({\cal R}^{-1} \vec v_E)_y 
=& 
v_E \left( 
-\sin(\varphi)[\left\{\sin^2(\Omega_E t)\cos(I_E)
+\cos^2(\Omega_E t)\right\}\sin(\omega_E t) 
\right.
\notag\\
& 
\left.
~~~+\left\{\sin(\Omega_E t)\cos(\Omega_E t)(1-\cos(I_E))\right\}
\cos(\omega_E t)] 
\right.
\notag\\
& 
\left.
~~~+\cos(\varphi)\sin(\Omega_E t)\sin(I_E) 
\right) , 
\nonumber\\ 
({\cal R}^{-1} \vec v_E)_z 
=& 
v_E \left(
\cos(\varphi)[\left\{\sin^2(\Omega_E t)\cos(I_E)+\cos^2(\Omega_E t)\right\}
\sin(\omega_E t) 
\right.
\notag\\
& 
\left.
~~~+\left\{\sin(\Omega_E t)\cos(\Omega_E t)(1-\cos(I_E))\right\}
\cos(\omega_E t)] 
\right.
\notag\\
& 
\left.
~~~+\sin(\varphi)\sin(\Omega_E t)\sin(I_E)
\right) .
\label{vE}
\end{align}

Let us investigate the time at which the phase shift becomes 
maximum each day. 
The expression of the above factor is linear in both 
$\sin(\omega_E t)$ and $\cos(\omega_E t)$, 
because it is linear in the velocity $\vec v_E$. 
Therefore, it is shown that $\partial(\Delta_{\mbox{CS}})/\partial t = 0$ 
is approximately equal to 
\begin{equation}
\tan(\omega_E t) = F(\Omega_E t, \varphi, I_E, \vec N_{I0}). 
\label{time}
\end{equation}
Here, $F$ is a lengthy function of $\Omega_E t$, 
$\varphi$, $I_E$ and $\vec N_{I0}$
 without including $\omega_E$ and $v_E$,  
and we have used $\Omega_E \ll \omega_E$ that implies 
$|\partial(\sin\Omega_E t)/\partial t| 
<< |\partial(\sin\omega_E t)/\partial t|$. 

Let us consider three cases that 
$\vec N_I$ is vertical upward, eastbound and northbound. 
The extremum condition Eq. $(\ref{time})$ 
becomes simpler and it implies that, 
independent of $\varphi$, 
the maximum phase difference occurs at 
6, 0 (and 12), and 18 hours of each day, 
when the normal vector to the interferometer is 
vertical, eastbound and northbound, 
respectively. 
This result agrees with numerical ones.

\subsection{Equatorial cases}
For its simplicity, Ref. \cite{OYA} 
made numerical calculations only for the equatorial case. 
Figures \ref{f2} and \ref{f3} show 
numerical results of time variations in the phase difference.  
For the northbound case, one-day variation is extremely small. 
Hence, the dashed curve for $\Delta_{\mbox{CS}}$ in Figure $\ref{f2}$ 
looks like a horizontal straight line. 

$\Delta_{\mbox{CS}}$ on the winter solstice day is the same as  
that on the summer solstice day. 
This is because the angles between Earth's axis 
and its orbital velocity on the two days agree with each other. 
On the other hand, 
$\Delta_{\mbox{CS}}$ on the spring equinox day and 
that on the autumn equinox day are the same with the opposite sign, 
since the relative angles on the days are too.

\subsection{Nonequatorial cases}
\begin{figure}[t]
\includegraphics[width=8cm]{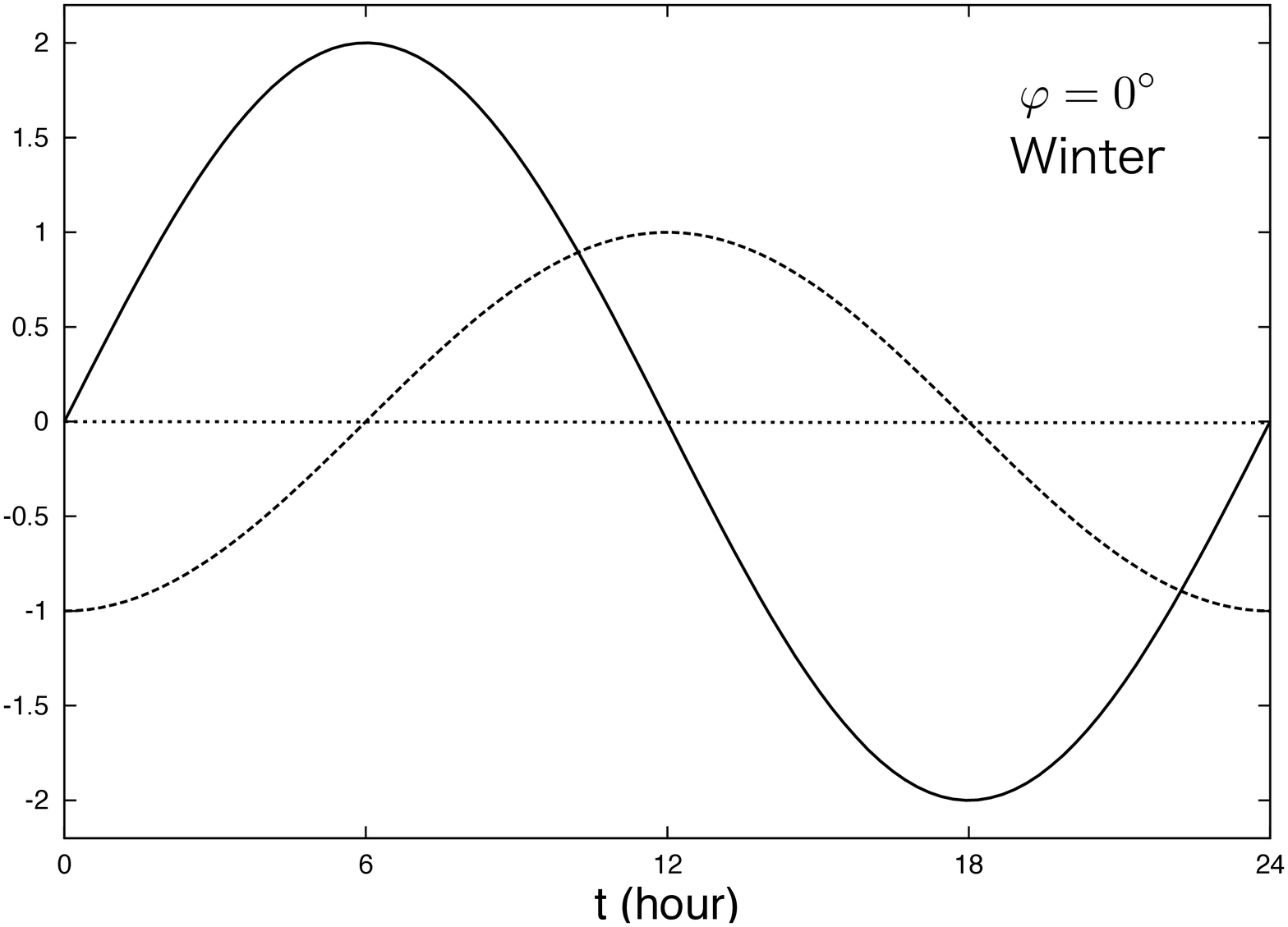}
\includegraphics[width=8cm]{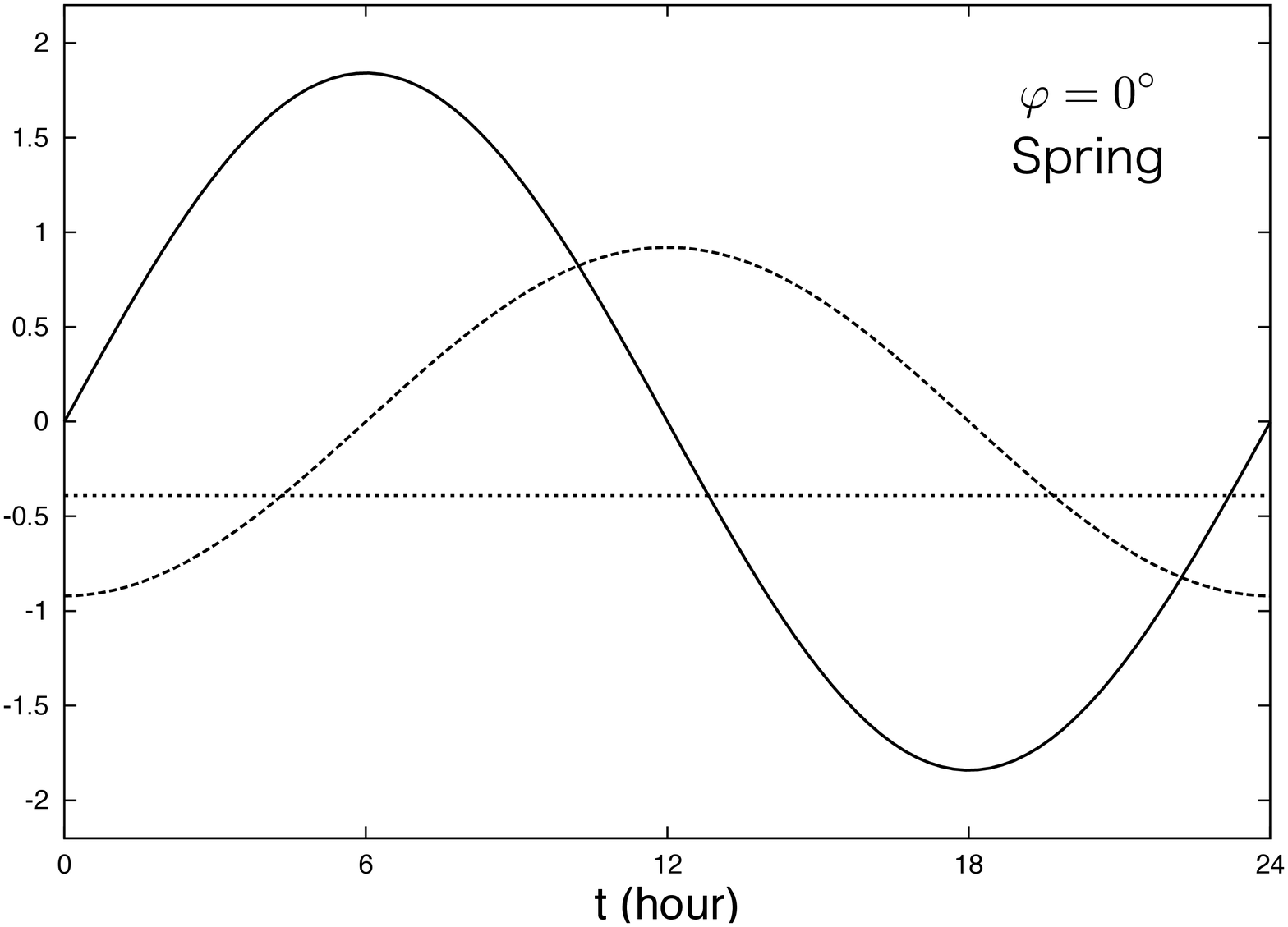}
\includegraphics[width=8cm]{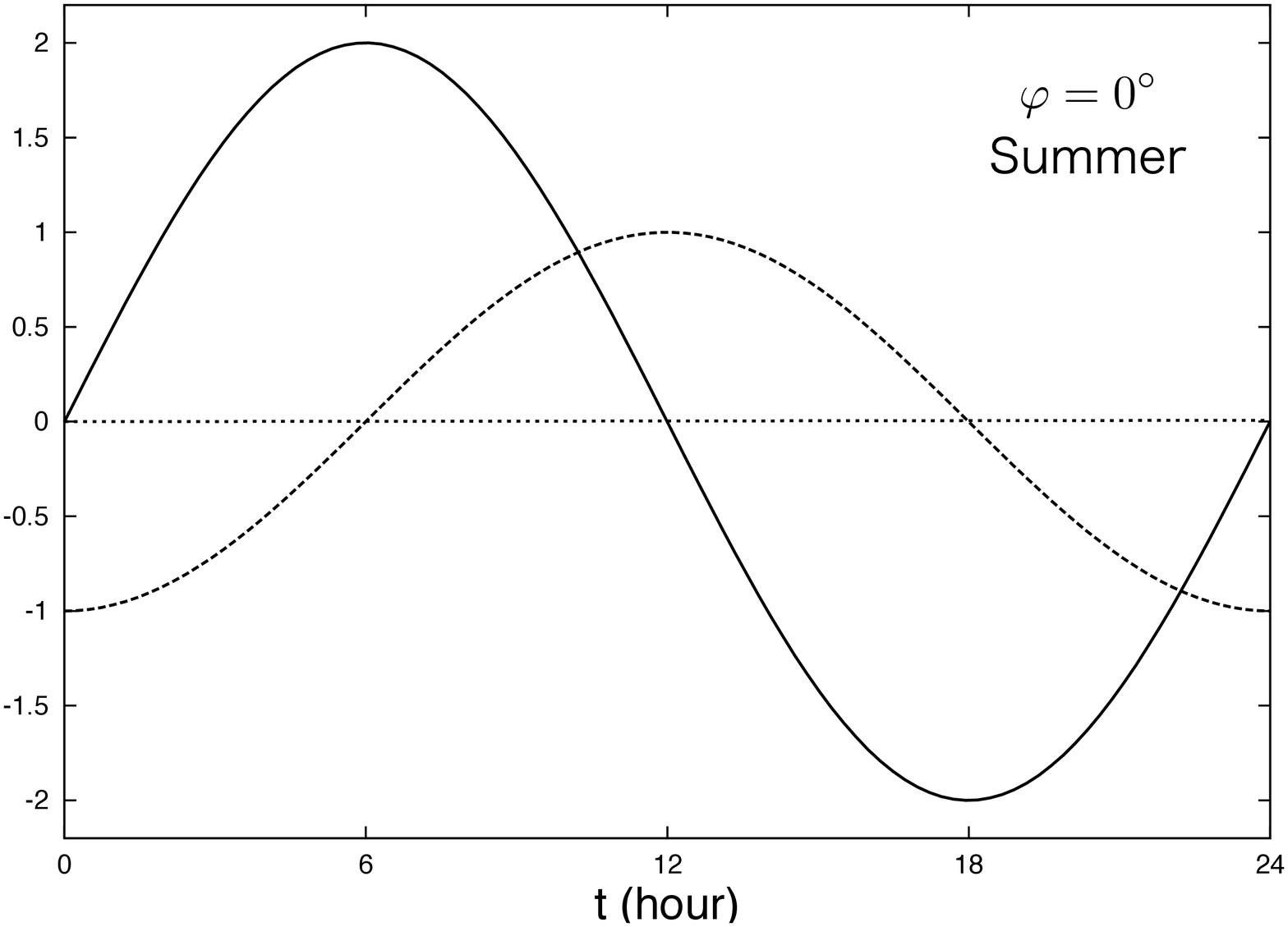}
\includegraphics[width=8cm]{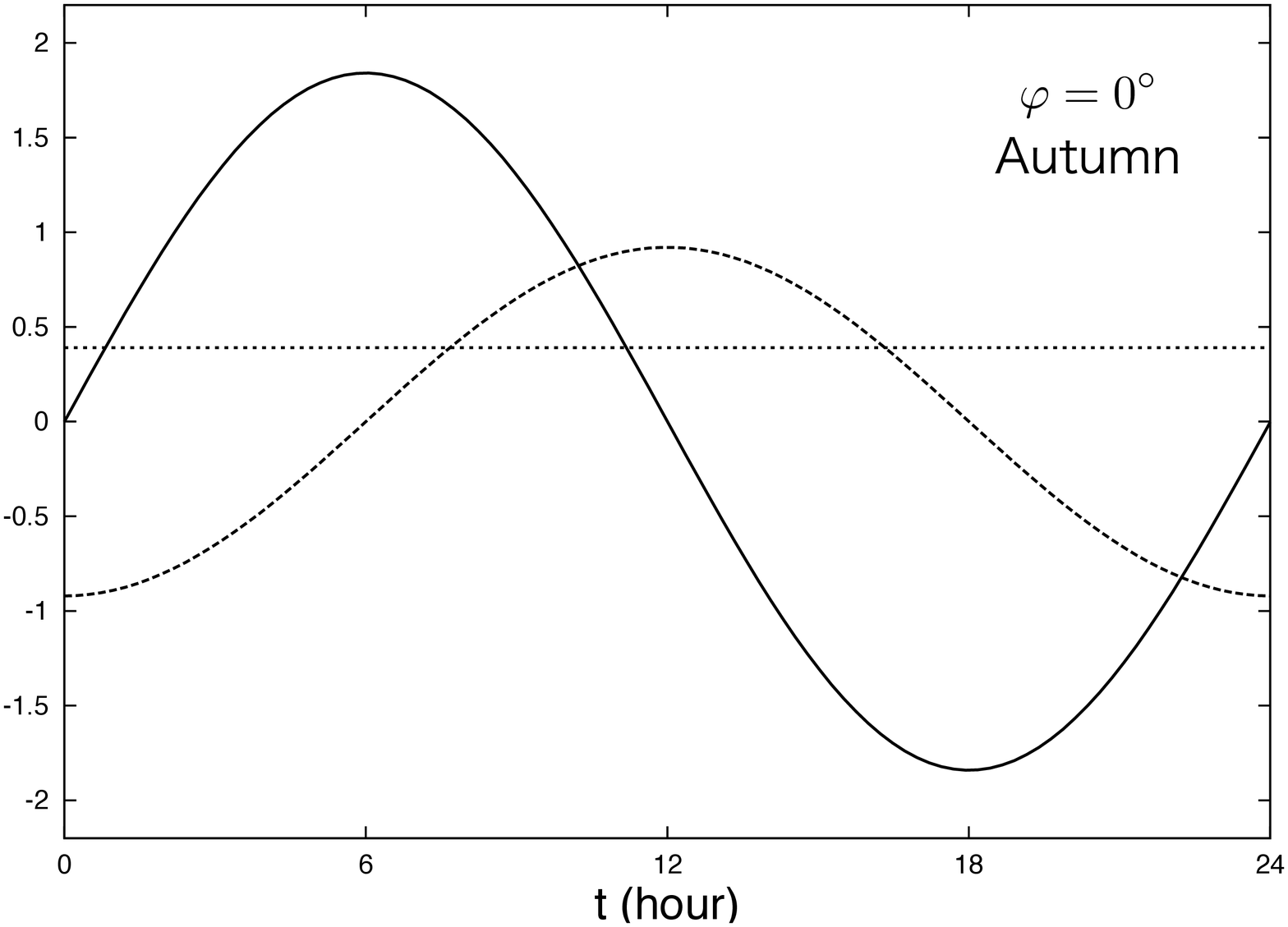}
\caption{ 
One-day variation in phase differences by CS effects 
on the quantum interferometer located 
on the equator of Earth (in local time). 
Top left: Winter solstice day. 
Top right: Vernal equinox day. 
Bottom left: Summer solstice day. 
Bottom right: Autumnal equinox day. 
The vertical axis denotes 
$\left[3(\vec v_E \cdot \vec n_E)\vec n_E -\vec v_E \right] 
\cdot\vec N_I$  
in Eq. (\ref{Delta-CS}) in the units of $v_E=1$. 
We consider three cases of the interferometer direction. 
The solid, dashed and dotted curves correspond to $\vec N_I$ 
that is vertical upward, eastbound and northbound, respectively. 
}
\label{f2}
\end{figure}

\begin{figure}[t]
\includegraphics[width=8cm]{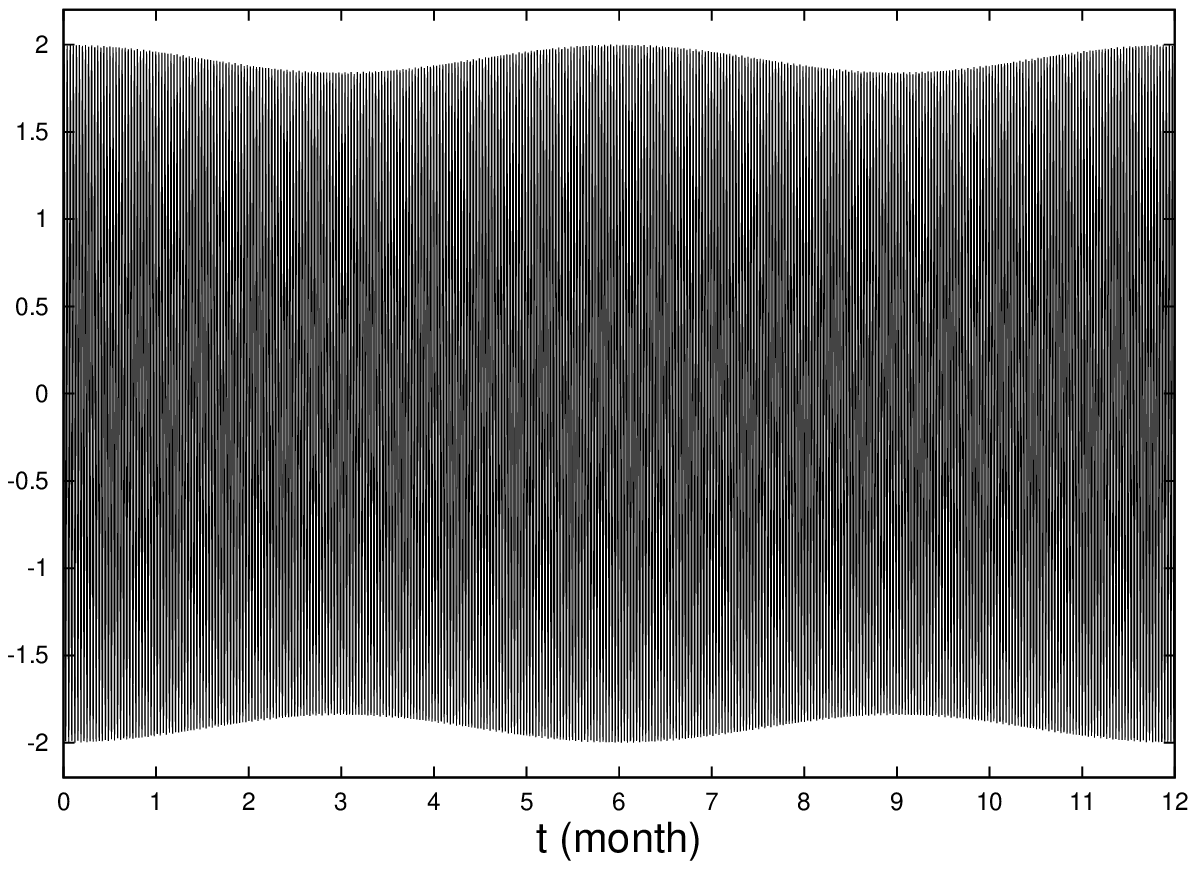}\\
\includegraphics[width=8cm]{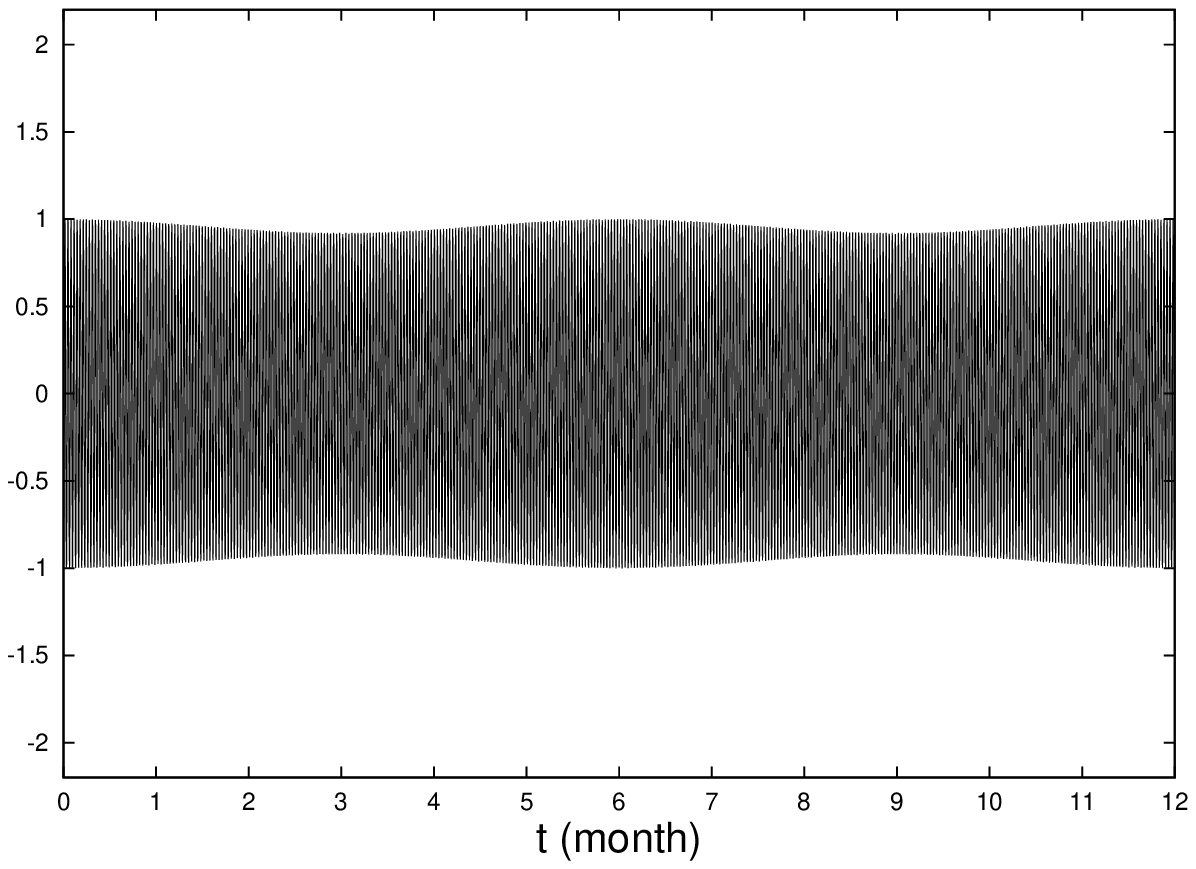}\\
\includegraphics[width=8cm]{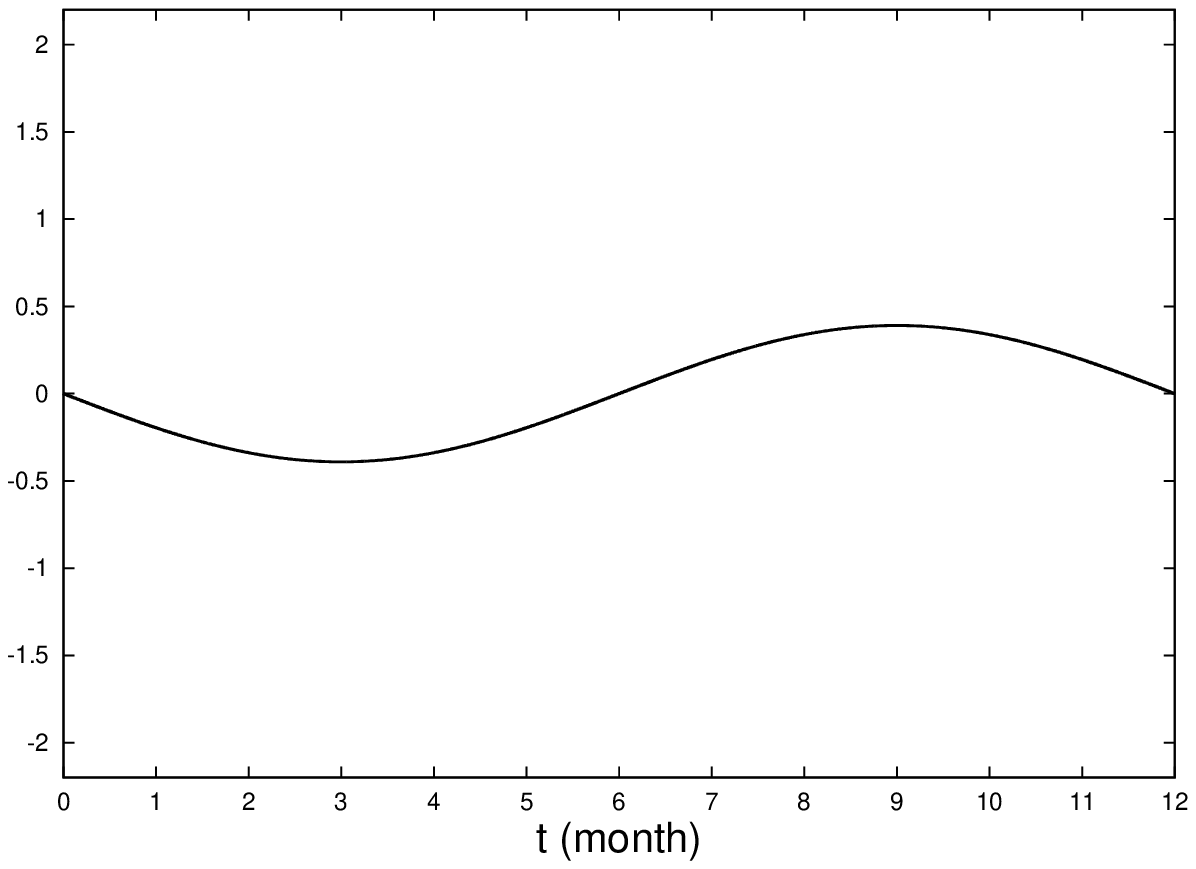}
\caption{ 
Seasonal variation in phase differences by CS effects 
on the same quantum interferometer in Fig. \ref{f2}. 
The winter solstice day is chosen as 0 month 
and the summer solstice 
corresponds to six months. 
Upper: Vertical $\vec N_I$ 
(corresponding to the solid curve in Fig. \ref{f2}). 
Middle: Eastbound $\vec N_I$ 
(corresponding to the dashed curve in Fig. \ref{f2}).
Bottom: Northbound $\vec N_I$ 
(corresponding to the dotted curve in Fig. \ref{f2}).
}
\label{f3}
\end{figure}

\begin{figure}[t]
\includegraphics[width=8cm]{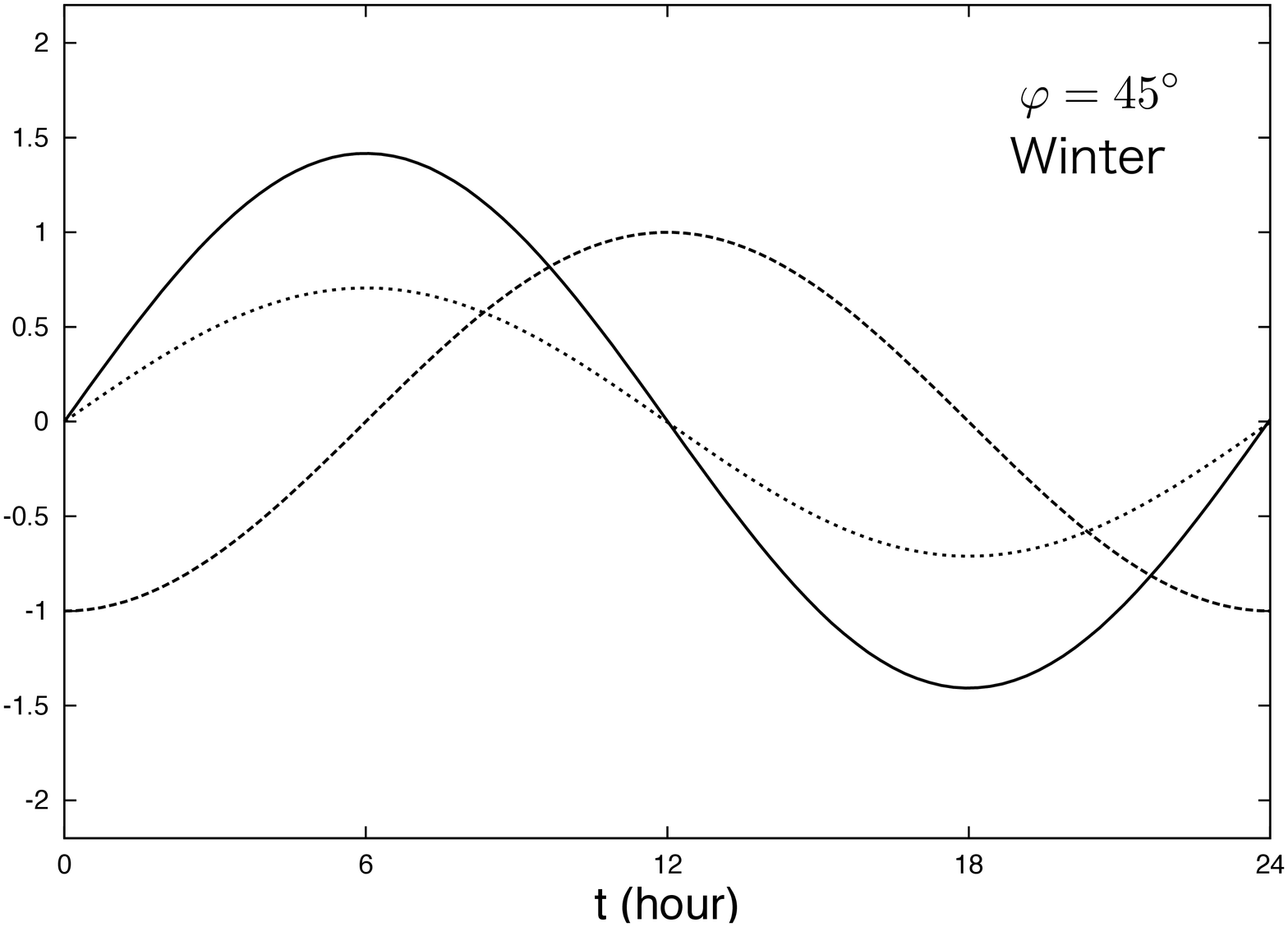}
\includegraphics[width=8cm]{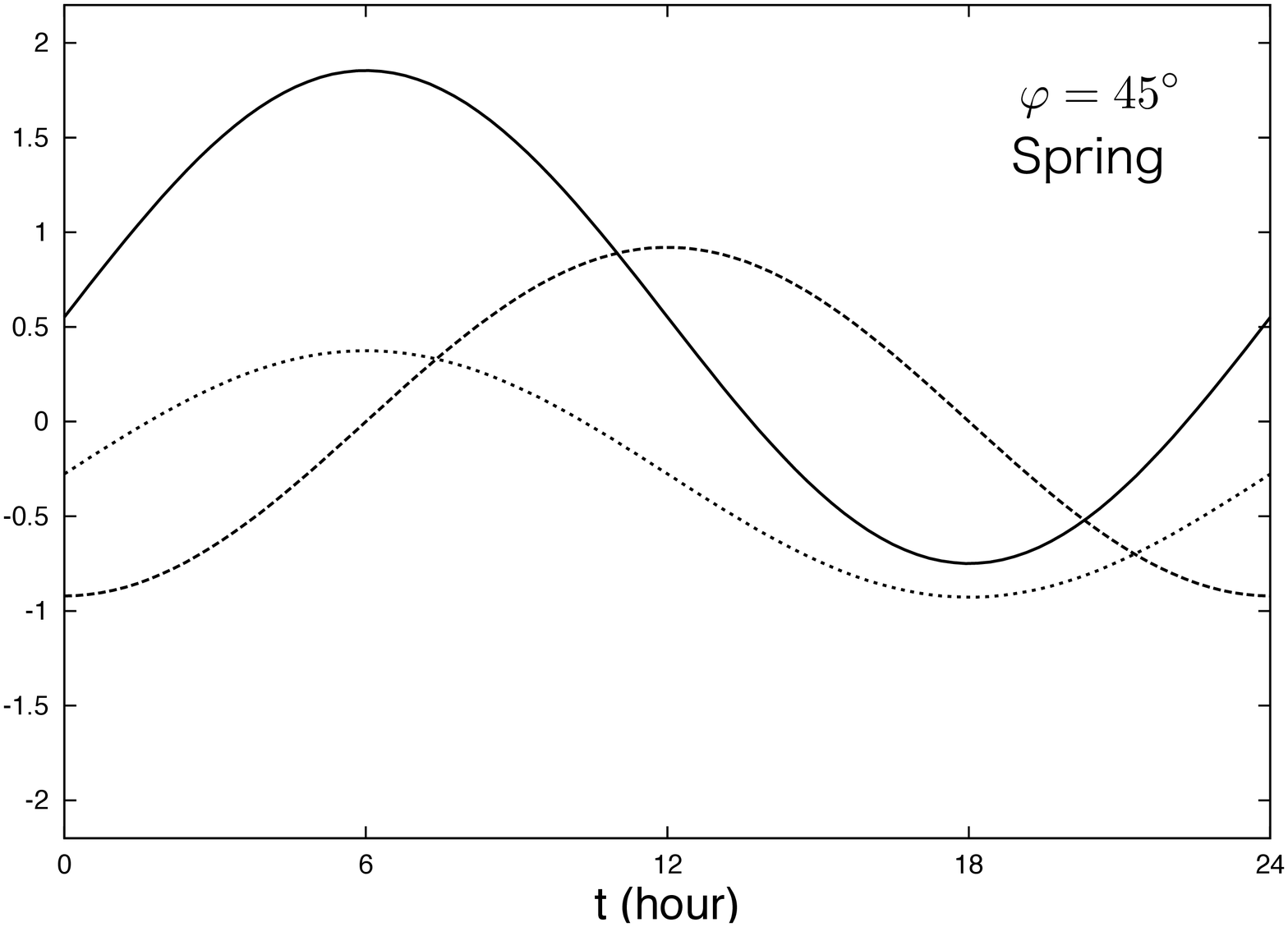}
\includegraphics[width=8cm]{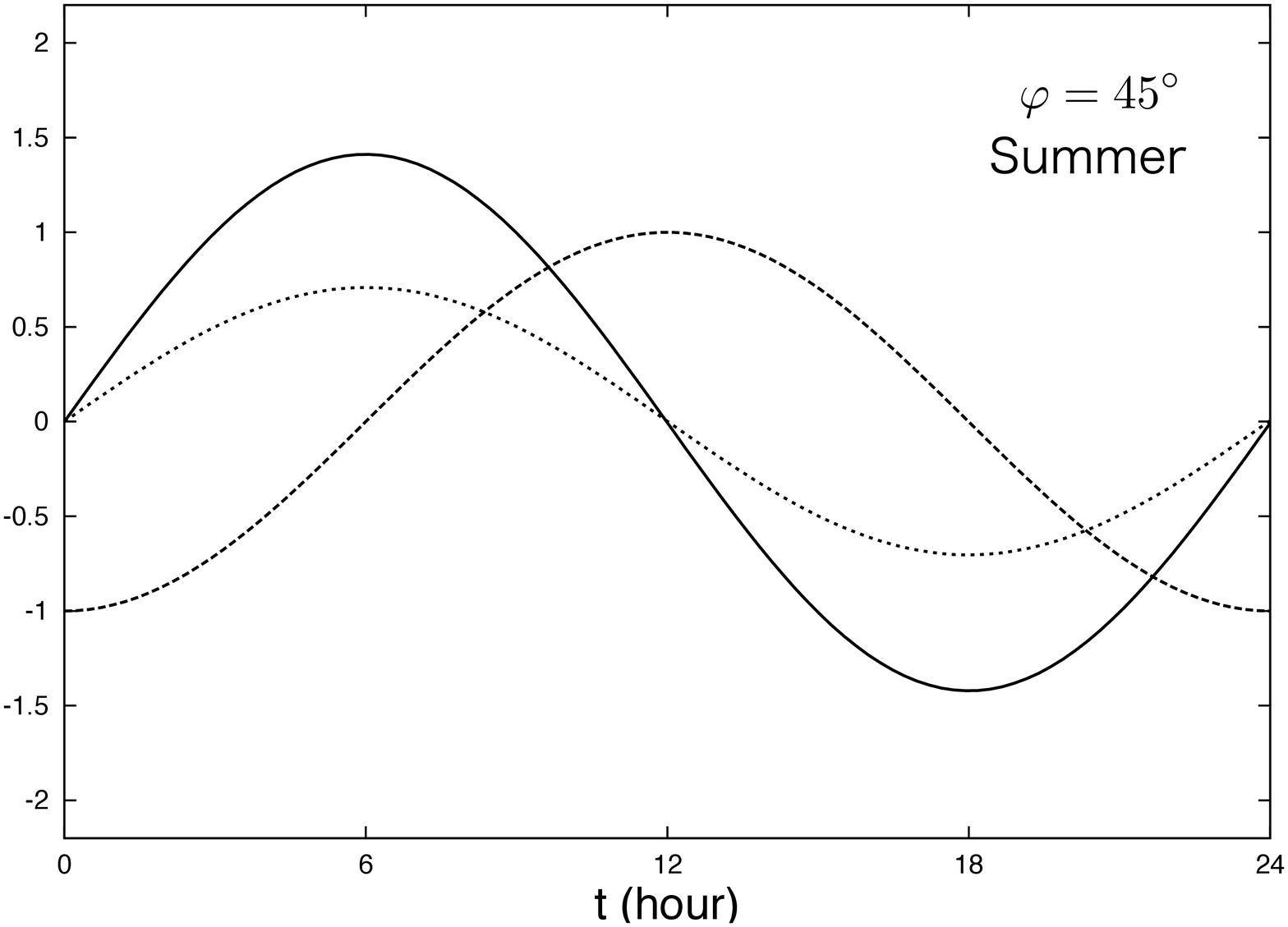}
\includegraphics[width=8cm]{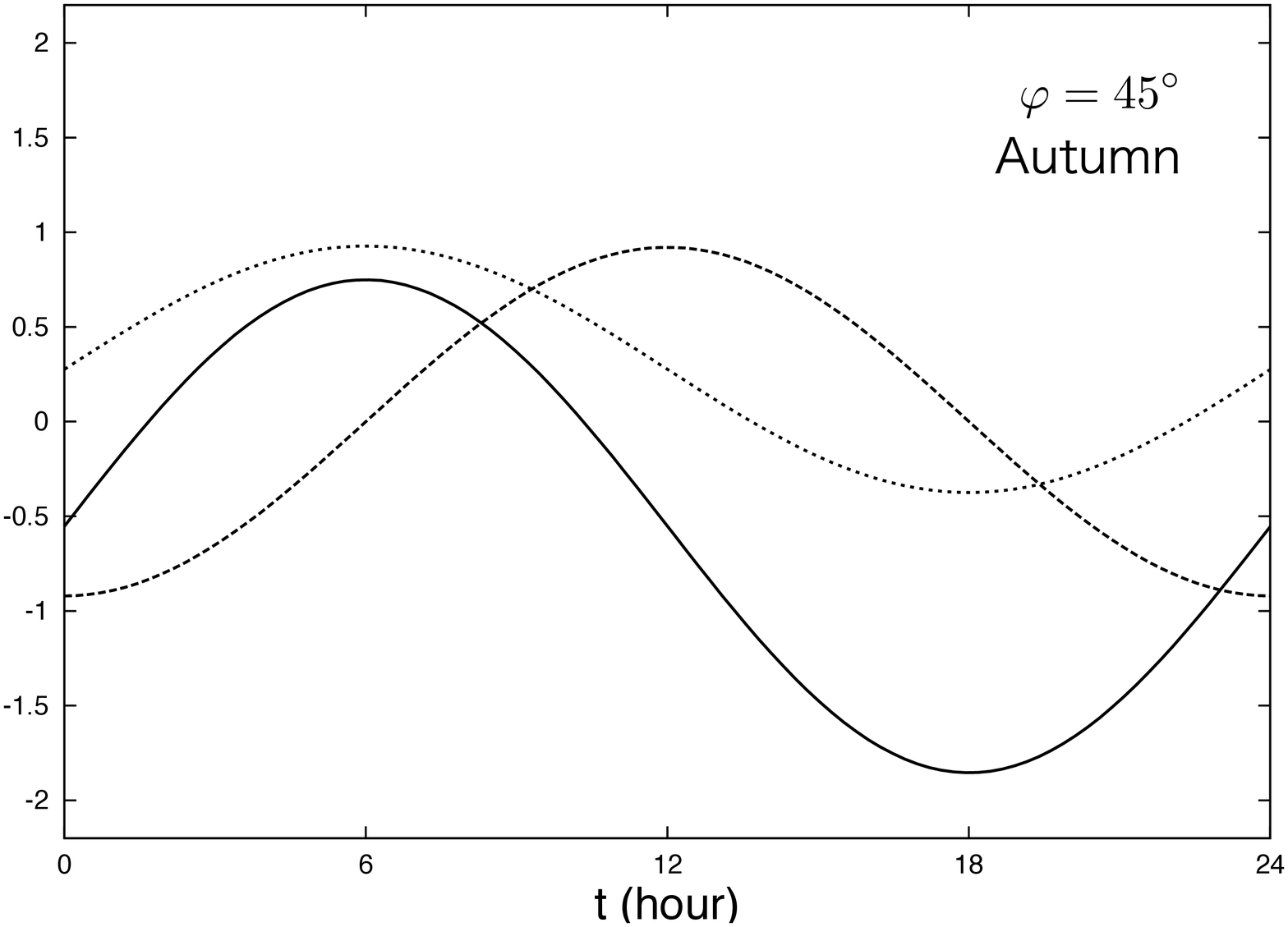}
\caption{ 
One-day variation at the latitude $\varphi = 45^{\circ}$, 
corresponding to Figure $\ref{f2}$ 
for $\varphi = 0^{\circ}$.  
Top left: Winter solstice day. 
Top right: Vernal equinox day. 
Bottom left: Summer solstice day. 
Bottom right: Autumnal equinox day. 
The solid, dashed and dotted curves 
correspond to $\vec N_I$ that is vertical upward, 
eastbound and northbound, respectively. 
}
\label{f4}
\end{figure}

For the case of $\varphi = 45^{\circ}$, 
Figure $\ref{f4}$ shows variations of the phase shift 
on the winter solstice day, the vernal equinox day, 
the summer solstice day and the autumn equinox day. 
For two cases as $\vec N_I$ being along the vertical direction 
and along the East one, 
time variation behaviors are slightly different from  
the equatorial cases in Figure $\ref{f2}$, 
whereas, for $\vec N_I$ pointing the North direction, 
they are significantly different from the equatorial case. 
Namely, the phase shift for the northbound $\vec N_I$ case 
depends more strongly on the latitude. 
Eqs. $(\ref{factor})$ and $(\ref{vE})$ suggest that 
$\Delta_{\mbox{CS}}$ is linear in both $\sin(\varphi)$ and $\cos(\varphi)$. 
For a small latitude difference $\delta\varphi$, therefore,  
the change in $\Delta_{\mbox{CS}}$ would be 
$O(\delta\varphi) \sim O(\sin\delta\varphi)$. 

\begin{figure}[t]
\includegraphics[width=8cm]{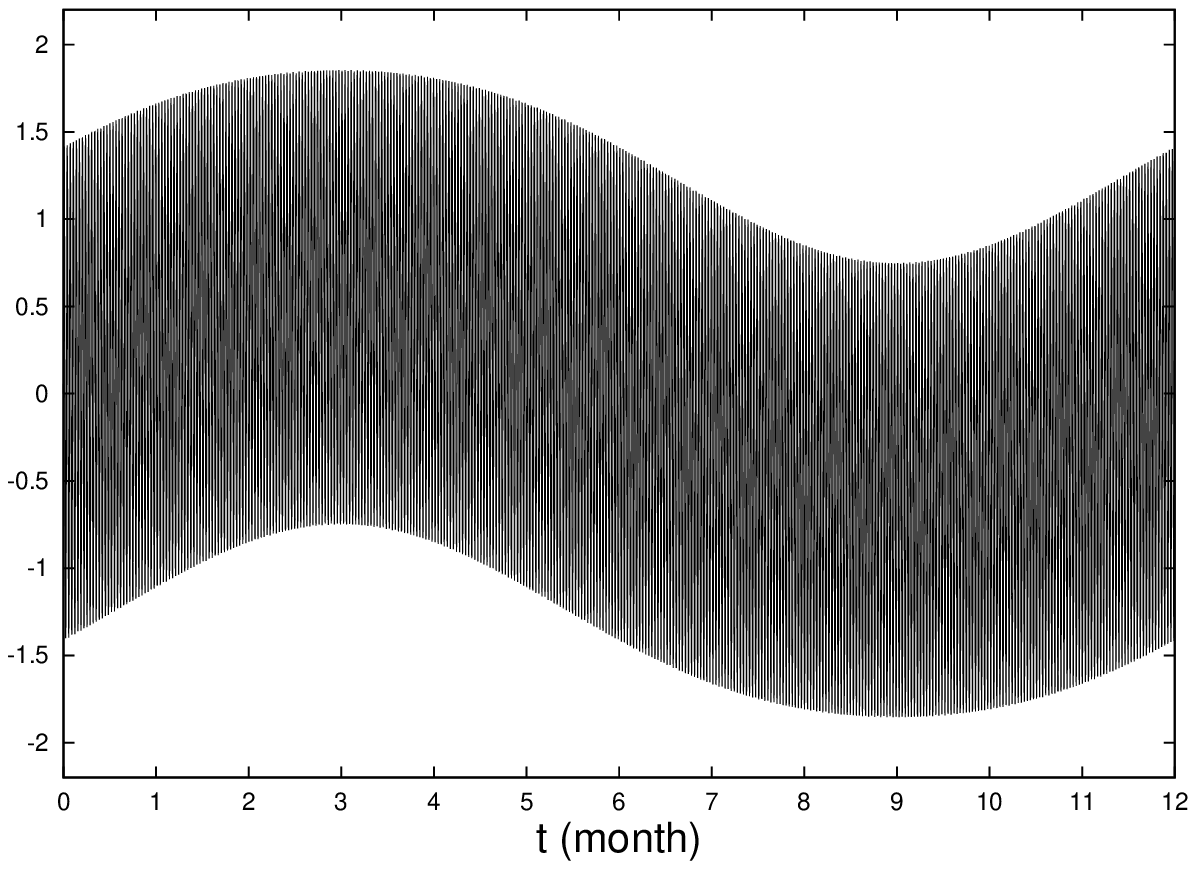}\\
\includegraphics[width=8cm]{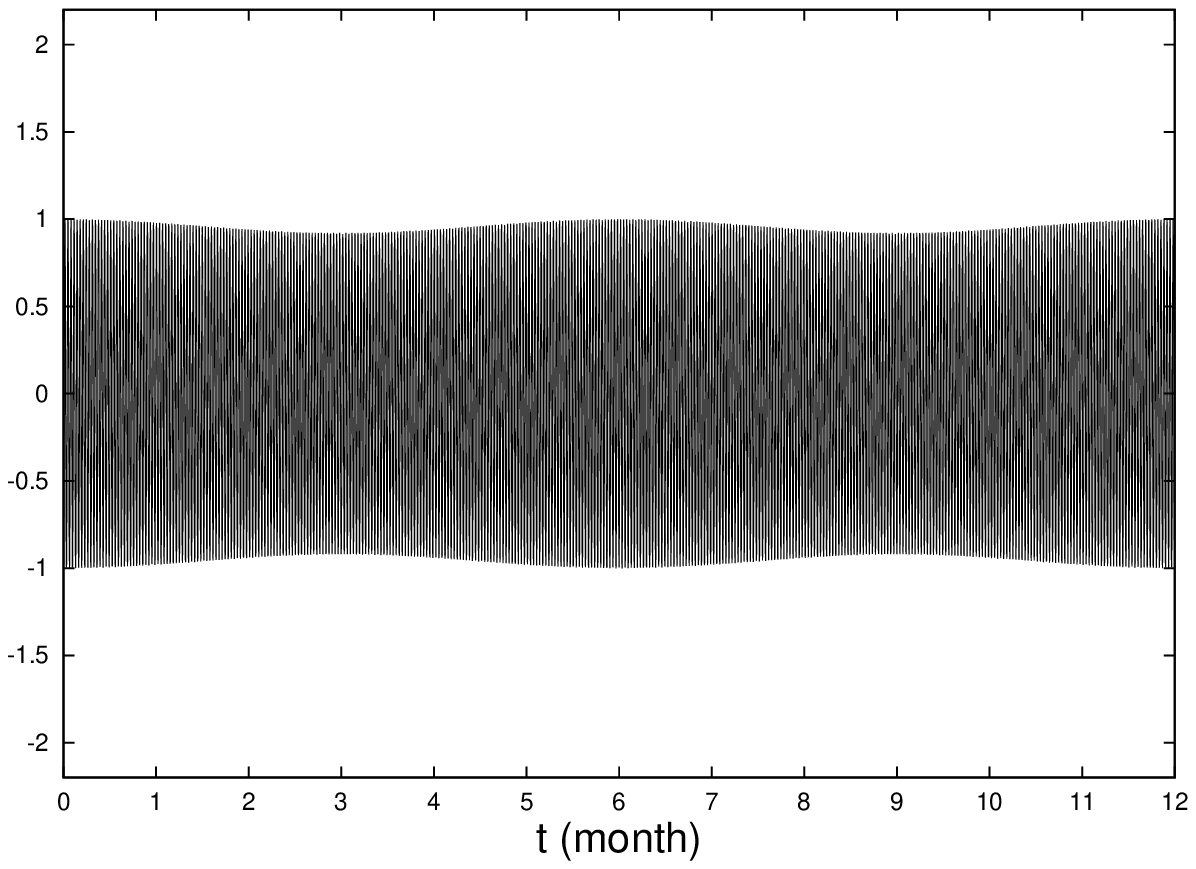}\\
\includegraphics[width=8cm]{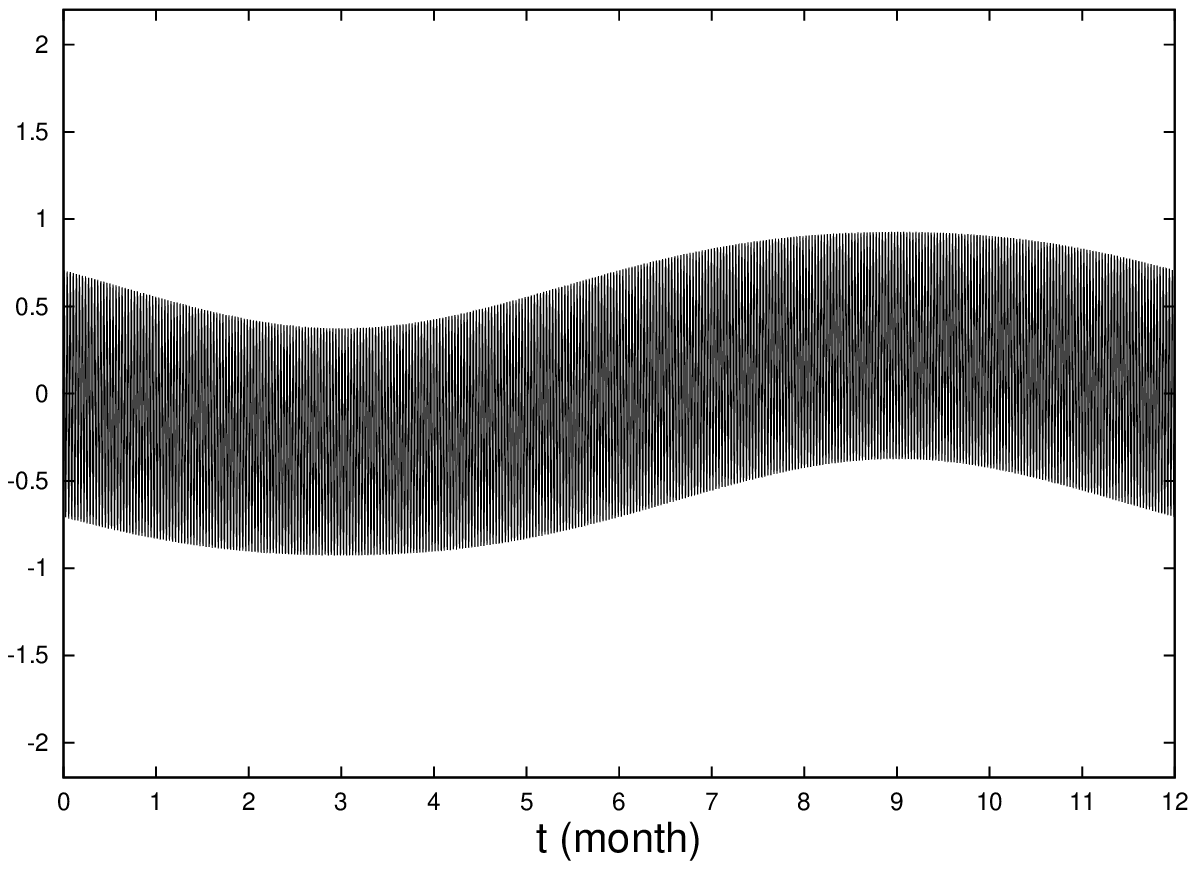}
\caption{ 
Seasonal variation at the latitude $\varphi = 45^{\circ}$ 
corresponding to Figure $\ref{f3}$. 
Upper: Vertical $\vec N_I$ 
(corresponding to the solid curve in Fig. \ref{f4}). 
Middle: Eastbound $\vec N_I$ 
(corresponding to the dashed curve in Fig. \ref{f4}).
Bottom: Northbound $\vec N_I$ 
(corresponding to the dotted curve in Fig. \ref{f4}).
}
\label{f5}
\end{figure}

The seasonal variation for $\varphi = 45^{\circ}$ 
is plotted in Figure $\ref{f5}$. 
Like the seasonal variation of the weather 
such as the maximum and minimum temperatures, 
the seasonal variation of $\Delta_{\mbox{CS}}$ 
is more strongly dependent on the latitude 
than the daily variation. 
On all the four days, 
the change of $\Delta_{\mbox{CS}}$ for the northbound case 
is smallest. 
More generally, Figure $\ref{f5}$ suggests that, 
the one-day variation for the northbound case is always 
smaller than those for the other two cases. 
Therefore, the northbound case might be preferred in 
testing a CS model by quantum interferometry. 

What is a role of the CS latitude effect in experiments? 
Without any anomaly in phase shifts measured by experiments, 
detailed theoretical templates of the time variations 
would not be needed. 
The latitude effect seems less important for this case. 
On the other hand, if some variation were marginally detected in the future 
(conceivably at a low signal-to-noise ratio),  
it would be difficult to distinguish the CS signal from 
other effects without taking account of the latitude effect. 
Comparing phase measurements at two (or more) latitudes 
would be helpful for improving the CS bound or 
distinguishing the CS signal from others. 
Namely, a signal-to-noise ratio could be increased 
by a combined analysis of phase measurements at different latitudes.

\section{Conclusion}
For investigating possible latitude effects of 
Chern-Simons gravity on quantum interference, 
we performed numerical calculations of time variation 
in the induced phase shifts for nonequatorial cases. 
At any latitude, 
the maximum phase shift might occur at 
6, 0 (and 12), and 18 hours (in local time) of each day, 
when the normal vector to the interferometer is 
vertical, eastbound and northbound, 
respectively.

If two identical interferometers were located at different latitudes, 
the difference between phase shifts that are measured 
at the same local time 
would be $O(\sin \delta\varphi)$ 
for a small latitude difference $\delta\varphi$. 
It might become maximally $\sim 20$ percents 
for $\delta\varphi \sim 10$ degrees, for instance. 

In the northbound case, 
the daily variation is very small. 
Therefore, the northbound case might be used for 
effectively improving the statistics 
by combining many data points for each day.

We would like to thank N. Yunes and S. Takeuchi  
for useful discussions. 
This work was supported in part (K.Y.) 
by Japan Society for the Promotion of Science, 
Grant-in-Aid for JSPS Fellows, No. 24108.


\begin{thebibliography}{99}
\bibitem{Refs}
L. A. Page, Phys. Rev. Lett. {\bf 35}, 543 (1975); 
J. Anandan, Phys. Rev. D {\bf 15}, 1448 (1977); 
M. Dresden and C. N. Yang, Phys. Rev. D {\bf 20}, 1846 (1979); 
J. J. Sakurai, Phys. Rev. D {\bf 21}, 2993 (1980); 
L. Parker, Phys. Rev. D {\bf 22}, 1922 (1980); 
Y. N. Obukhov, Phys. Rev. Lett. {\bf 86}, 192 (2001); 
X. Huang and L. Parker, Phys. Rev. D {\bf 79}, 024020 (2009); 
K. Konno and R. Takahashi, 
Phys. Rev. D {\bf 85}, 061502(R) (2012). 


\bibitem{COW}
R. Colella, A. W. Overhauser, and S. A. Werner, 
Phys. Rev. Lett. {\bf 34}, 1472 (1975).

\bibitem{Sakurai-Book}
J. J. Sakurai, {\it Modern Quantum Mechanics}  
(The Benjamin/Cummings Publishing, Menlo Park, CA, 1985), Chap. 2. 


\bibitem{Zych}
M. Zych, F. Costa, I. Pikovski, C. Brukner, 
Nature Communications {\bf 2}, 505 (2011). 

\bibitem{Hogan}
C. J. Hogan, Phys. Rev. D {\bf 85}, 064007 (2012). 

\bibitem{Polchinski}
J. Polchinski, {\it Superstring Theory and Beyond String Theory} vol.2 
(Cambridge University Press, Cambridge, UK, 1998). 
\bibitem{Ashtekar}
A. Ashtekar, A.P. Balachandran, and S. Jo, 
Int. J. Mod. Phys. A {\bf 4}, 1493 (1989).  

\bibitem{Taveras}
V. Taveras and N. Yunes, 
Phys. Rev. D {\bf 78}, 064070 (2008); 
S. Mercuri and V. Taveras,
Phys. Rev. D {\bf 80}, 104007 (2009).

\bibitem{AY1}
S. Alexander and N. Yunes, Phys. Rev. Lett. {\bf 99}, 241101 (2007). 
\bibitem{AY2}
S. Alexander and N. Yunes, Phys. Rev. D {\bf 75}, 124022 (2007). 

\bibitem{AY2009} 
S. Alexander and N. Yunes, Phys. Rep. {\bf 480}, 1 (2009). 

\bibitem{Ali-Haimoud}. 
Y. Ali-Haimoud and Y. Chen, 
Phys. Rev. D {\bf 84}, 124033 (2011).

\bibitem{Dyda} 
S. Dyda, E. E. Flanagan and M. Kamionkowski, 
arXiv:1208.4871. 

\bibitem{Nandi}
K. K. Nandi, I. R. Kizirgulov, O. V. Mikolaychuk, 
N. P. Mikolaychuk, and A. A. Potapov, 
Phys. Rev. D {\bf 79}, 083006 (2009). 

\bibitem{OYA}
H. Okawara, K. Yamada, H. Asada, 
Phys. Rev. Lett. {\bf 109}, 231101 (2012). 

\bibitem{Jackiw}
R. Jackiw and S. Y. Pi, Phys. Rev. D {\bf 68}, 104012 (2003).
\bibitem{Guarrera}
D. Guarrera and A. J. Hariton, Phys. Rev. D {\bf 76}, 044011 (2007).  

\bibitem{Grumiller}
D. Grumiller and N. Yunes, 
Phys. Rev. D {\bf 77}, 044015 (2008).

\bibitem{Sopuerta}
N. Yunes and C. F. Sopuerta, 
Phys. Rev. D {\bf 77}, 064007 (2008). 

\bibitem{rotation}
N. Yunes and F. Pretorius, 
Phys. Rev. D {\bf 79}, 084043 (2009);
K. Yagi, N. Yunes and T. Tanaka, 
Phys. Rev. D {\bf 86}, 044037 (2012);
K. Yagi, L. C. Stein, N. Yunes, and T. Tanaka, 
arXiv:1302.1918.

\bibitem{Will}
C. M. Will, {\it Theory and Experiment in Gravitational Physics} 
(Cambridge University Press, Cambridge, UK, 1993).

\bibitem{AYcorr}
$\vec \nabla \times \vec g = \vec \Omega$ 
is calculated in Ref. \cite{AY1, AY2}. 
Note that the coefficient $-1$ in the right hand side of 
Eq. (15) in Ref. \cite{AY1} and Eq. (61) in Ref. \cite{AY2} 
should read $2$. 
(N. Yunes, private communications 2011)

\bibitem{AY3}
References \cite{AY1, AY2} adopt the geometrical units 
($G=c=1$), with which quantum experimentalists 
may not be familiar. 
Hence, $G$ and $c$ are kept in this paper. 

\bibitem{LL}L. D. Landau and E. M. Lifshitz, {\it The Classical Theory 
of Fields} (Pergamon, New York, 1962). 

\bibitem{Chambers} 
R. G. Chambers, Phys. Rev. Lett. {\bf 5}, 3 (1960).  


\bibitem{Kuroiwa}
J. Kuroiwa, M. Kasai, and T. Futamase, 
Phys. Lett. A {\bf 182}, 330 (1993). 
\bibitem{Wajima}
S. Wajima, M. Kasai, T. Futamase, 
Phys. Rev. D {\bf 55}, 1964 (1997). 

\bibitem{Book} 
H. Rauch, and S. A. Werner, 
{\it Neutron Interferometry} (Oxford University Press, Oxford, 2000). 

\bibitem{GPB}
C. W. F. Everitt et al. Phys. Rev. Lett. {\bf 106}, 221101 (2011). 

\bibitem{Smith} 
T. L. Smith, A. L. Erickcek, R. R. Caldwell, and M. Kamionkowski, 
Phys. Rev. D {\bf 77}, 024015 (2008). 

\bibitem{Werner} 
S. A. Werner, and A. G. Klein, 
J. Phys. A: Math. Theor. {\bf 43}, 354006 (2010). 



\bibitem{Seki}
Y. Seki, H. Funahashi, M. Kitaguchi, M. Hino, Y. Otake, 
K. Taketani, and H. M. Shimizu, 
J. Phys. Soc. J. {\bf 79}, 124201 (2010).


\bibitem{Pushin}
D. A. Pushin, M. G. Huber, M. Arif, and D. G. Cory, 
Phys. Rev. Lett. {\bf 107}, 150401 (2011). 


\end{thebibliography}
\end{document}